\documentstyle[12pt]{article}
\begin{document}
\newcommand{\newc}{\newcommand}
\newc{\ra}{\rightarrow} 
\newc{\lra}{\leftrightarro} 
\newc{\beq}{\begin{equation}} 
\newc{\eeq}{\end{equation}} 
\newc{\barr}{\begin{eqnarray}} 
\newc{\err}{\end{eqnarray}}
\newc{\eps}{\epsilon}
 \def\gappeq{\mathrel{\rlap {\raise.5ex\hbox{$>$}}
{\lower.5ex\hbox{$\sim$}}}}
\def\lappeq{\mathrel{\rlap{\raise.5ex\hbox{$<$}}
{\lower.5ex\hbox{$\sim$}}}}
\def\l{\label}
\def\r{\ref}
\newcommand{\gl}{\lambda}
\newcommand{\PR}{{\it Phys. Rev. }}
\newcommand{\PL}{{\it Phys. Lett. }}
\newcommand{\NP}{{\it Nucl. Phys. }}

$ $\hfill hep-ph/9608323

~~\hfill IOA.02/96

~~\hfill NTUA 55-96\\
\vspace{1cm}

\begin{center}
{\bf
Duality Constraints on  Supersymmetric Unified Models\\
and Radiative Symmetry Breaking$^\dagger$
}

\vspace*{.5cm}
{\bf 
G.K. Leontaris$^{(a)(1)}$ and N.D. Tracas$^{(b)(c)(2)}$
} \\
\vspace*{.5cm}
{\it $^{(a)}$Theoretical Physics Division, Ioannina University}\\
{\it GR-451 10 Ioannina, Greece}\\
\vspace*{.3cm}
{\it $^{(b)}$Physics Department, National Technical University}\\
{\it GR-157 80 Zografou, Athens, Greece}\\
\vspace*{.3cm}
{\it $^{(c)}$Theory Division, CERN}\\
{\it CH-1211 Geneva 23, Switzerland}
\end{center}

\begin{abstract}
Motivated from  unified models with string origin, we 
analyse the constraints from duality invariance on  effective
supergravity models with intermediate gauge symmetry. 
Requiring vanishing vacuum energy and invariance of the 
superpotential couplings, the modular weights are subject
to various constraints. Further, the intermediate gauge symmetry
breaking scale $M_U$ 
is  related to the values of modular weights of the matter and 
higgs fields. For certain regions of values of the latter, $M_U$ can
be close to the conventional unification scale of the minimal
supersymmetric standard  model. We also examine particular
examples where the intermediate gauge symmetry breaks down
to the standard gauge group radiatively.  
\end{abstract}

\thispagestyle{empty}
\vfill
\noindent IOA 02/96\\
\noindent NTUA 55-96\\
\noindent August 1996

\vspace{.5cm}
\hrule
\vspace{.3cm}
{\small
\noindent
$^\dagger$Work partially supported by C.E.C. Projects SC1-CT91-0729 and 
CHRX-CT93-0319.\\
$^{(1)}$leonta@cc.uoi.gr\\
$^{(2)}$ntrac@central.ntua.gr

\newpage

If one adopts the idea for unification  of all forces,  the minimal 
supersymmetric standard model (MSSM) is considered as the most natural 
extension of the standard model (SM) of strong and electroweak interactions,
since its spectrum allows the three gauge couplings to meet at an energy of
${\cal O}(10^{16}GeV)$
\cite{.}.
 Besides, unified supersymmetric models 
solve successfully the hierarchy problem
\cite{susy}
of their corresponding 
non-supersymmetric versions. Yet, the MSSM  has many arbitrary parameters   
and by no means can be considered as the ultimate theory of elementary 
particles. The only road beyond the MSSM which looks promising these
days is $N=1$ supergravity
\cite{sg}
coupled to matter and gauge fields.
However, $N=1$ supergravity still contains a lot of arbitrariness. One  is
free to choose the chiral multiplets provided they transform
consistently under the chosen gauge group, while Yukawa couplings in
the superpotential are also arbitrary. Nowadays, string theory appears as
the only serious candidate which could predict all the above arbitrary
parameters. On the other hand, strings can have as a limit an effective
$N=1$ supergravity theory. In addition, the gauge group of most of the
string constructions, contrary to what was usually assumed in the old
Grand Unified approach
\cite{ross},
is predicted to have a product structure, rather
than being a single gauge group. Thus, if string theory really plays 
a role in particle physics, then one is left with an effective theory 
with the following summarized general characteristics.
There is an effective unification scale, namely the string scale 
$M_{str}$, where all couplings -- up to threshold corrections 
-- attain a common value.  At this point one is left with an effective
$N=1$ supergravity theory while the  gauge group structure is of
the form $G=\prod_nG_n$, which usually contains an `observable' and a
`hidden' part. In general, the observable part  has a rank larger than
that of the MSSM symmetry. In such a case, we may say that $G$ is an 
Intermediate Gauge Symmetry (IGS) which breaks down to the SM - gauge
group at an intermediate scale $M_X$, usually some two orders of
magnitude below the string scale. Moreover, string symmetries
give rather strong constraints on many of the parameters of the
effective field theory model. For example, the kinetic term 
appears to have a certain structure, while the effective Lagrangian
is in general invariant under certain duality symmetries which
act on the space of the moduli
\cite{dic}.
 The superpotential and the
Yukawa couplings is also subject to similar constraints.
In the present work, we would like to present an analysis
of the modular invariance constraints in effective models arising
from the string. In particular, we have in mind models with
 an IGS scale which   often appear in 
string constructions. We will examine the following issues.
Using the constraints of modular invariance we will determine
the properties of the effective potential in IGS models. 
Requiring also zero cosmological constant, we will correlate 
the vacuum expectation values of the higgs fields breaking
the IGS group, with the modular weights. We will finally present
a particular example where gauge symmetry may break radiatively.

In $N=1$ supergravity one introduces a real gauge invariant K\"ahler 
function whose general form is
\cite{FKZ}
\beq
G(z,\bar z) = {\cal K}(z,\bar z) + log{\mid {\cal W}(z)\mid}^2
\eeq
where ${\cal K}(z,\bar z)$ is the K\"ahler potential whose second derivatives
determine the kinetic terms for the various fields in the chiral supermultiplets
(we are using the standard supergravity mass units). 
${\cal W}$ is 
the superpotential which is a holomorphic function of the chiral
superfields.
Denoting $z=(\Phi,Q)$, where $\Phi$
stands for the dilaton field S and other moduli 
$T_i$ while Q for the chiral superfields, we may expand the K\"ahler potential
as follows
\beq
{\cal K}(\Phi,\bar{\Phi},Q,\bar Q)=
K (\Phi,\bar \Phi) +
Z_{\bar ij} (\Phi,\bar\Phi)\bar Q_{\bar i}e^{2V}Q_j+\cdots   \label{2}
\eeq
At the tree level ${ K}(\Phi,\bar{\Phi})$ is written  
\beq
K_t(\Phi,\bar{\Phi})=-log(S+\bar S)+K_0(T,\bar T)  \label{3}
\eeq
Higher terms are proportional to inverse powers of $(S+\bar{S})$,
\beq
{ K}(\Phi,\bar{\Phi})=K_t(\Phi,\bar{\Phi})+
\frac{1}{8\pi^2}\frac{K^{(1)}_{\bar ij}}{S+\bar S}
+\frac{1}{(8\pi^2)^2} \frac{K_{\bar ij}^{(2)}}{(S+\bar S)^{2}}+\cdots
 \label{4a}
\eeq
 while the
 kinetic energy matrix assumes a similar expansion
\cite{KL,FKZ}
\beq
Z_{\bar ij}=Z_{\bar ij}^0(T,\bar T)+
\frac{1}{8\pi^2}\frac{Z^{(1)}_{\bar ij}}{S+\bar S}
+\frac{1}{(8\pi^2)^2} \frac{Z_{\bar ij}^{(2)}}{(S+\bar S)^{2}}+\cdots
 \label{Zexp}
\eeq
In the above expansions, only the combination $S+\bar{S}$ of the dilaton
field appears, as a Peccei-Quinn (PQ) symmetry holds to all orders of
perturbation theory.
 
The superpotential ${\cal W}(z)$ is a holomorphic
function of the chiral superfields $Q_i$ and at the tree level is given by
\beq
{\cal W}({\Phi},Q) = \frac 13\lambda_{ijk}({\Phi})Q_iQ_jQ_k +
 \frac 12 \mu_{ij}(\Phi)Q_iQ_j + \cdots  \label{sp}
\eeq
where $\{ \cdots \}$ stand for possible non-renormalisable
contributions. Terms bilinear in the fields $Q_i$  refer in fact
to an effective higgs  mixing term
\cite{mu1,mu2}.
 Perturbative effects 
may allow dilaton contributions to the superpotential of the form 
$\propto e^{-8\pi^2S}$, thus breaking the original PQ
 symmetry which allowed only $S+\bar S$ dilaton combinations to appear. 

In the following we will assume that the tree  level  K\"ahler potential
 $ K_0 (T,\bar T)$  can take the following general factorizable form
\cite{EW,MC}:
\beq
{ K_0}(T,\bar T)=-\Sigma_nh_nlog(T_{n}+\bar T_{n}) \label{6}
\eeq
which implies the following form for the gravitino mass
\begin{equation}
m_{3/2}(z,\bar z)=
\frac{\mid {\cal W}(z)\mid}{\left(\prod_n(T_n+\bar{T}_n)^{h_n} 
(S+\bar{S})\right)^{1/2}} 
\end{equation}
Now under the modular symmetries, the moduli transform as
\beq
T\rightarrow \frac{aT -\imath b}{\imath cT+d} \label{7}
\eeq
where $a,b,c,d$ constitute the entries of the $SL(2,{\bf Z})$ group elements 
with $a,b,c,d\in {\bf Z}$ and $ad-bc=1$. Next, applying the  K\"ahler 
transformation,
\begin{eqnarray}
{\cal K} &\rightarrow&{\cal K}'={\cal K} +J(T)+J^*(\bar T) \\
{\cal W} &\rightarrow& {\cal W}'= {\cal W} exp\{-J(T)\} \label{8}
\end{eqnarray}
Eqs(\ref{6},\ref{7}) imply that $J(T)$ has the specific form
\beq
J(T)=\sum_nh_nlog(\imath  c_nT_n+d_n)\label{9}
\eeq
Following the same procedure, we may obtain the transformation properties 
of the tree level matrix $Z_{i\bar j}^{(0)}(T,\bar T)$ in the 
expansion of Eq(\ref{Zexp}). 
 $Z_{i\bar j}^{(0)}$ is given by the formula
\cite{AGN,IL}
\beq
Z_{i\bar j}^0=\delta_{i\bar j}\prod_n(T_{n}+\bar T_n)^{-q^n_i}\label{10}
\eeq
where the exponents are in general rational numbers.

Applying the SL(2,{\bf Z}) transformation, Eq(\ref{7}), to the tree 
level mass  term $Z_{i\bar j}^0Q_i\bar Q_{\bar j}$  we conclude 
that  the matter fields should transform as follows:
\begin{eqnarray}
Q_i&\rightarrow  
&\delta_{ij}Q_i\prod_nt_n^{-q_i^n}\label{11}
\end{eqnarray}
where, to simplify the subsequent formulae, we have introduced the notation
            \[t_n=\imath c_nT_n+d_n, 
                   \] 
The obtained formulae in Eqs(\ref{9},\ref{11}) 
may give further restrictions to the transformation  properties
 of the superpotential. Thus, the K\"ahler transformations Eq(\ref{8}) and 
the $J(T)$ form in
Eq(\ref{9}) imply that the perturbative superpotential is transformed
as follows
\beq
{\cal W}\rightarrow\prod_nt_n^{-h_n}
{\cal W}\label{spt}
\eeq
We consider in the following the various terms in the superpotential separately.  
In fact we are interested in the two types of terms of the perturbative tree
level superpotential exhibited in Eq(\ref{sp}). Thus, the transformation property
Eq(\ref{spt}) together with that of the fields $Q_i$ in Eq(\ref{11}), imply that
the $\mu$ parameter is transformed as follows 
\beq
\mu '_{ij}=\mu_{ij}\prod_lt_l^{-h_l}\prod_m t_m^{q_i^m}
\prod_nt_n^{q_j^n} \label{mup}
\eeq
A similar expression is expected to hold for the $\gl_{ijk}$ parameters of 
the superpotential.

Thus far, the above procedure gives us no further constraints on the
superpotential terms,  Yukawa couplings and mass parameters. However,
we may impose the constraint that the latter remain invariant under 
transformations
implied by the symmetries of the string. Therefore, we assume  the tree 
level Yukawa  couplings and the $\mu$ parameter to be invariant (up to 
a moduli - independent phase) scalar  functions under the action of the 
modular transformations
{\footnote
{ Actually in $Z_2\times Z_2$ orbifold construction, the Yukawa couplings
are constants, while in Calabi-Yau manifolds they approach a constant
value in the large volume limit of the moduli they depend on
\cite{FKZ}.
}.
Thus, if we demand  invariance of the $\mu$ - 
term -- ignoring for simplicity the possible existence of a $T$ - independent
 phase -- we obtain the following relation of modular weights
\beq
\prod_lt_l^{h_l}=\prod_mt_m^{q_{2}^m}\prod_nt_n^{q_1^n}\label{C_1}
\eeq

A similar reasoning for the case of
 the Yukawa mass term for the chiral fields
$Q_i$ leads to the following general condition for the modular weights,
\beq
\prod_l t_l^{h_l}=\prod_m t_m^{q_i^m}\prod_n t_n^{q_j^n}
\prod_r t_r^{q_k^r}  \label{C_2}
\eeq 

The Eqs(\ref{C_1},\ref{C_2}) are obtained by the simultaneous action 
of the SL(2,Z) modular invariance constraints and the invariance of
 the superpotential parameters of the effective field theory model. 
In fact  they provide specific relations among the modular weights 
whose role is  decisive for the initial conditions of the scalar fields, 
as can be seen from the form of the tree level scalar matrix $Z_{i\bar j}^0$. 

The parameters of the theory may further be restricted if one imposes the 
cosmological constant constraint. In order for our procedure to be more 
transparent, let us simplify the subsequent analysis, assuming that there 
is a flat direction where $T_1=T_2=\cdots =T_N\equiv T$ which means that
the potential depends on a single modulus T. In this case setting $h_n=3/N$ 
we can simply  write, $K_0=-3\log (T+\bar T),$ to ensure zero vacuum energy. 
Returning now to the constraint Eq(\ref{C_1}), assuming that this
 is for a higgs  mixing term $\mu H_1H_2$, we may obtain the following     simple relation
 between the modular  weights for the two standard model higgs fields,
\beq
q_1+q_{2}=h \label{q_1}
\eeq
where $h= N\cdot h_n =3$ for the simple example presented here.
Furthermore, if the relation Eq(\ref{C_2}) has been obtained, say, for a 
trilinear coupling of the up-type quark mass matrix $Q H_2 u^c$, a similar
constraint may emerge for the corresponding modular weights too, i.e., 
$q_Q+q_{2}+q_{u^c}=h$. These constraints can be easily generalised
in the case of models with couplings transforming covariantly under
the modular symmetries.

Let us see how this type of constraints may give information about the 
scalar masses at the unification scale.
Consider the simple case of untwisted sectors where modular weights are 
integers. From relation Eq(\ref{q_1}) we may obtain for example $q_1=1,
q_{ 2}=2 $ or $q_{2}=1, q_1=2 $ leading to two distinct cases for 
the initial conditions for the  scalar masses 
$m_{H_1}^2, m_{H_2}^2$. 
In the MSSM for example, such a distinction 
is welcome; in fact, it implies non - universal boundary conditions for
the two higgs doublets which is essential in particular for cases with 
large $tan\beta$, i.e., with approximately equal top and bottom Yukawa
couplings at the unification scale.
Thus, from the two sets of $q_{1,2}$ values above, one may choose the 
phenomenologically viable case which finally drives  the one $(mass)^2$
parameter negative at a low scale so that radiative symmetry breaking of 
the $SU(2)\times U(1)$ occurs.
If higgs particles arise from twisted sectors then $q_{1}, q_{2}$ can
be any rational number and many solutions can exist even under the
apparently restrictive relation Eq(\ref{q_1}). 

The above discussion presumes that the $SU(3)\times SU(2)\times U(1)$ 
symmetry arises  at the string scale.
It is rather difficult however to obtain correct values of the low 
energy coupling constants with the ordinary matter fields in the 
massless spectrum.
Indeed, it is well known that using only the MSSM spectrum, unification
of the gauge  couplings occurs naturally at $M_U\sim 10^{16} GeV$
\cite{.},
 i.e. 
almost two orders lower than the string scale. This is rather suggestive
for the existence of an IGS.
Models with IGS have appeared in a string 
context
\cite{FERCON,STM,KD}.
 Thus, in the following we wish to extend our previous 
analysis in these latter cases. 
In fact, our main motivation for this analysis are 
string derived models 
based on $SU(4)\times O(4)$, $SU(5)\times U(1)$ 
and  $SU(3)^3$ symmetries. Thus  in what follows, we will assume 
that there  is at least one pair of higgs fields, $H_{1,2}$, having 
the required group properties, and  obtaining large vacuum 
expectation values (vevs) which break the intermediate gauge
group down to the MSSM symmetry. For example, in the case of 
$SU(4)\times O(4)\sim SU(4)\times  SU(2)_L\times SU(2)_R$ these 
may be $H_1= (4,1,2)$ and $H_2 =(\bar{4},1,2)$.
In  $SU(5)\times U(1)$ these are $5,\bar{5}$ and $10,\bar{10}$. 
(A particular application  will be presented in the subsequent 
analysis.) We will also perform our computation in the context of our
previous simplification, i.e.,  considering only one modulus $T$ and 
the dilaton field $S$.

With respect to the fields $z_I = (z_S,z_0,z_1,z_2)\equiv 
(S, T, H_1, H_2)$, the scalar potential $V(z)$ is given by
\beq
V = e^{G(z)}\left(G_I G^{-1}_{I\bar{J}}G_{\bar{J}} -3\right) 
+ \mid D\mid^2
\label{pt}
\eeq
where $\mid D\mid^2$ represents the contribution of the  D-terms in the 
potential.
Also, with $G_I$, we denote the derivatives of $G$ with respect to the
fields $z_I$, i.e.,
\barr
G_I & \equiv 
\frac 1{\cal W}{\cal D}_I{\cal W}
\err
where ${\cal D}_I{\cal W}=\partial_I{\cal W}+{\cal W}\partial_I{\cal K} $ 
is the K\"ahler derivative. Thus, with respect to the moduli $T$, 
we have for example
\beq
(T+\bar T)\partial_T{\cal K}= -h-q_{i}Z_i(T,\bar T)H_i\bar{H_i}
\eeq
and analogously for $\partial_{\bar T}{\bar{\cal W}}$.
In the above basis $z_I$, the K\"ahler metric has the following block
diagonal form
\barr
G_{I\bar{J}}\equiv {\cal K}_{I\bar{J}}
=\left(\begin{array}{cc}{\cal K}_{S\bar{S}}& 0\\
                        0 &{\cal K}_{i\bar{j}}\\
       \end{array}\right)
\label{gij}
\err 
where the subscripts denote differentiation with respect to the fields $z_I$
while ${\cal K}_{i\bar{j}}$ is a $3\otimes 3$ matrix with the indices $i,j$
taking the values $0,1,2$ for the fields $T,H_1,H_2$ respectively.

In order to calculate the potential Eq(\ref{pt}), we need the inverse K\"ahler
metric $G_{I{\bar J}}^{-1}$. In particular, ${\cal K}_{i\bar j}^{-1}$ is given by
\barr
{\cal K}_{i\bar j}^{-1}=\frac{1}{\rho^2}
\left(\begin{array}{ccc}
   1   & \bar{\eta}_1                       &\bar{ \eta}_2\\
\eta_1 & \tau^{q_1}\rho^2+\mid \eta_1\mid^2 &\eta_1\bar{\eta}_2\\
\eta_2 & \bar{\eta}_1\eta_2                 &\tau^{q_2}\rho^2+
                                                     \mid\eta_2\mid^2\\
\end{array}\right)
\err
where, we have introduced the convenient notation
\beq
\tau^2\rho^2 \equiv Q^2 = h+q_1\frac{H_1\bar{H}_1}{\tau^{q_1}}+
                            q_2 \frac{H_2\bar{H}_2}{\tau^{q_2}}
\label{qu}
\eeq
with $\tau = T+\bar T$, while  
\beq
\eta_i = \frac{q_iH_i}{T+\bar{T}}\equiv \frac{q_iH_i}{\tau}\quad,\qquad i=1,2
\eeq
The  scalar potential can now be obtained by computing the quantities
$G^IG_I\equiv G^IG_{I\bar J}^{-1}G^{\bar J}$. 
In oder now to examine in detail the properties of the
scalar potential, we need the specific knowledge of  the superpotential
couplings and in particular that part related to the higgs sector.
However,  for illustrative purposes let us ignore  derivative terms 
and collect only terms independent of $\mid {\cal W}\mid^2$.
In terms of the unrenormalized field vevs 
 $\upsilon_i=<H_i>$ we obtain, at the
minimum,                            
\beq
V_0=e^{<G>}\{3-(\frac{h^2}{Q^2}+\frac{\upsilon^2_1}{\tau^{q_1}}+ 
\frac{\upsilon^2_2}{\tau^{q_2}})\}\label{dov}
\eeq
where $Q^2$ is obtained from Eq(\ref{qu}) by substituting 
$\upsilon_i=<H_i>$.
Eq(\ref{dov}) determines the two vevs $\upsilon_{1,2}$ of
the IGS breaking higgs fields.
Assuming that the potential at the minimum is zero and equal 
renormalised vevs (so that the flatness of the
effective potential is ensured), we can express the vev as
 a function of $h$ and the sum $q_1+q_2\equiv q$. For $h$
close to 3 and the additional constraint $q=3$, we can get
a sensible result of vev$\sim 10^{-2}M_{str}$. The 
terms dropped out of the potential could allow for a 
wider range of $q$ and $h$ giving vev in the desired region.
This will be discussed in a future work.

Up to now, we have analysed how one can determine the higgs vevs
which break the IGS
and we have expressed their magnitudes as a function of the modular
weights. 
However, we have not yet referred to
the mechanism triggering the IGS breaking down to the standard
model. In the following, we would like to examine the possibility 
of breaking the IGS radiatively, pretty much the same way as this
happens in the MSSM
\cite{radbr}.
Now, the question we would like to ask is if a similar
phenomenon may occur in the case of an intermediate symmetry.
In a string unified model with IGS, as is the case we are examining
here, there are mainly two large scales involved. The first is the string
scale $M_{str}\sim 5\times 10^{17}GeV$, where one is left with a string
spectrum having transformation properties under the intermediate
gauge group ($SU(4)\times O(4), SU(5)\times U(1)$, etc). The
second large scale is the one where the IGS
breaks down to the standard model and it is usually assumed
to be approximately two orders of magnitude less than the string
scale $M_{str}$, i.e., it usually coincides with the scale $M_U$.
The more interesting cases arise in particular IGS models where 
a much lower intermediate scale is possible, so that negative
radiative corrections can grow up enough to turn a higgs 
$(mass)^2$ - parameter negative. This will be the case for the
 example we will present in the subsequent example. 
 
{}From the K\"ahler potential one may  obtain through  
the $Z(T,\bar T)$
matrix  soft mass parameters for the  IGS higgs
multiplets discussed in detail  previously. Their
magnitude is controlled by the supersymmetry breaking scale, so
their initial values may well be ${\cal O}(\le 1TeV)$. The question
which now arises is whether one of these $(mass)^2$ - parameters 
turns  negative at the right scale so that the IGS 
 breaks down radiatively. Now, in the case of MSSM,
there are two basic ingredients whose role is decisive: {\it i)} the huge
gap  of the $M_U, M_Z$ scales which allows radiative corrections to grow
up and, {\it ii)} the large top-Yukawa coupling. Instead, here  we first
note  that the gap between the two scales $M_{str}$ and the conventional
supersymmetric unification scale $M_U$ is rather 
short, $M_{str}/ M_U\sim 10^{2}$, and at first sight, it looks rather
unlikely that radiative corrections can do the job. Second, a large 
Yukawa coupling is needed to mimic the role of the top-Yukawa one,
in the low energy case. In order to see if the scenario of 
Radiative Intermediate Symmetry  Breaking (RISB) can occur, we
will take as an example the $SU(4)\times O(4)
\sim SU(4)\times SU(2)\times SU(2)$ model. Here, left 
and right handed fermions (including the right handed neutrino)
are accommodated in the  $F = (4,2,1)\;,\;\bar{F}=(\bar 4,1,2)$
representations respectively.  The SM symmetry breaking 
occurs due to the presence of the two standard doublet higgs fields  
which are found in the $h =(1,2,2)$ representation of the
original symmetry of the model 
(The decomposition of the $h$ under the 
$SU(3)\times SU(2)\times U(1)$ gauge group is 
$h(1,2,2)\ra h_u(1,2,\frac 12 ) + h_d(1,2,-\frac {1}{2})$.)
The $SU(4)\times SU(2)_R \ra SU(3)\times U(1)$ symmetry breaking
is realized at a high scale, 
\cite{422rge}
with the introduction
of a higgs pair belonging to  $H + \bar{H}= (4,1,2) + (\bar 4,1,2)$
representations. Sextets fields $D=(6,1,1)$ appear also in the model.
The gauge invariant tree level superpotential
which is of relevance to our discussion here is
\cite{AL}
 \barr
\cal W &=&\lambda_1F_L\bar F_Rh+\lambda_2\bar F_RH\phi_i
+\lambda_3HDD+\lambda_4\bar H\bar HD+\lambda_5\phi_0hh 
\label{sp4}
\err
{}From the terms shown in Eq(\ref{sp4}), one can easily figure out that
an essential role in the evolution of the soft mass terms 
{\footnote
{ In this letter, we will not discuss the origin of supersymmetry
breaking. For an attempt in the context of this model see however
\cite{mur}.}
of the
higgs fields is played by the couplings $\lambda_3 H H D$ and 
$\lambda_4 \bar{H} \bar{H} D$.  According to our discussion, these
couplings should be chosen sufficiently large if they are supposed
to play the role of the top Yukawa coupling in the minimal case.
Such a requirement is welcome here since the same couplings determine
 the masses of the higgs colour triplets living in  $D, H, \bar{H}$ 
which should acquire large masses in order to avoid fast proton decay.
In fact, under the SM gauge group the decomposition of the
above representations gives
$D(6,1,1)\ra D_3 +\bar{D}_3$ and  $H(4,1,2)\ra d_H+e_H+H^0$ 
and similarly for   $\bar{H}(\bar{4},1,2)\ra \bar{d}_{\bar{H}}+
e_{\bar{H}}+\bar{H}^0$. Now, down quark type colour triplets
will acquire large masses
\cite{AL}
 after the symmetry breaking,
proportional to $\lambda_{3,4}$, while $e_H,e_{\bar{H}}$
will be eaten by the higgs mechanism.  

To make clear how the couplings $\lambda_{3,4}$ are involved
in the evolution of the soft mass parameters 
for the neutral higgs components, let us write
the corresponding renormalization group equations. We
simplify the analysis by ignoring all other Yukawas.
We obtain (renaming $ H^0 = H_1$ and $ \bar{H}^0= H_2 $)
\cite{422rge}
\begin{eqnarray}
\frac{d\gl_3}{dt}&=&\frac{1}{8\pi^2}\gl_3
\left( 4\gl_3^2+\gl_4^2-\frac{25}{4}g_4^2-\frac{3}{2}g_R^2\right)\nonumber\\
\frac{d\gl_4}{dt}&=&\frac{1}{8\pi^2}\gl_4
\left( \gl_3^2+4\gl_4^2-\frac{25}{4}g_4^2-\frac{3}{2}g_R^2\right)\nonumber\\
\frac{dm^2_{H_1}}{dt}&=&\frac{1}{8\pi^2}
\left[\gl^2_3\left(6m^2_{H_1}+3m^2_D\right)-
       \frac{15}{2}g^2_4M^2_4-3g^2_RM^2_R\right]\label{rg4}\\
\frac{dm^2_{H_2}}{dt}&=&\frac{1}{8\pi^2}
\left[\gl^2_4\left(6m^2_{H_2}+3m^2_D\right)-
       \frac{15}{2}g^2_4M^2_4-3g^2_RM^2_R\right]\nonumber\\
\frac{dm^2_D}{dt}&=&\frac{1}{8\pi^2}
\left[\gl^2_3\left(2m^2_D+4m^2_{H_1}\right)+
      \gl^2_4\left(2m^2_D+4m^2_{H_2}\right)
      -10g^2_4M^2_4\right]\nonumber
\end{eqnarray}
where $M_i$ stands for the gaugino mass of the corresponding group factor.
In the following we perform a numerical investigation of the above
set of equations in order to find whether it is possible to obtain
a negative $(mass)^2$.  First we determine the initial values of the soft 
higgs mass parameters  from the potential.  Taking the
derivatives of  the potential, the soft masses can in general
be of the form
\cite{IL}
\beq
m_{soft}^2 = m_{3/2}^2 + V_0 + {\rm modular\;\; weight \;\; 
dependent\;\; terms}
\eeq
where $V_0$ is essentially the cosmological constant.
Thus, in our case, after rescaling to obtain correct normalized fields, 
while assuming zero cosmological constant, we get 
\beq
m^2_{H_i}=m^2_{3/2}(1 + q_i)\quad,\qquad i=1,2\label{hm}
\eeq

Obviously, the initial conditions of the two higgs fields depend 
crucially on the modular weights $q_{1,2}$ which are in general not
 equal to each other. In the case of untwisted fields, 
they are integer numbers otherwise  they can be any rational number.
 In any case, according to our assumptions 
$q_i$'s should satisfy the constraint  (\ref{q_1}). In order to
present an illustrative example, we take $h=3, q_1=11/4, q_2=1/4$ and
$m_{3/2}= 100GeV$. We assume further, large initial values for
the Yukawa couplings $\lambda_{3,4}\sim {\cal O}(1)$. 
In Figure (2a) the two higgs mass - parameters
are depicted as a function of the scale $\log_{10}M$. It can be seen that
one of them turns negative at a  scale $M_X\sim 2\times 10^{15}GeV$,
 not far from the conventional unification scale $M_U$. All other
soft squared mass parameters are positive at that scale.
In the model under consideration, the IGS breaking scale $M_X$ can 
be even lower, without running into phenomenological troubles  as 
the gauge bosons mediating fast proton decay are absent in this model. 
Another example is presented in Figure (2b). Here we consider integer
modular weights, $q_1=2, q_2=1$, while again we take $m_{3/2}=100$GeV.
As expected, $m_{H_2}^2$ is driven now negative at a lower scale.
The maximum IGS breaking scale is of course obtained when $q_1=3,
q_2=0$ so that the initial mass parameters have the maximum gap,
$m_{H_2}/m_{H_1}=1/\sqrt{3}$. For comparison, we show 
the $m_{H_2}$ - plot for the  three selective $(q_1,q_2)$ pairs
 in Figure 3. In all the above figures, we choose for convenience
to plot $m_{H_i}$'s instead of $m_{H_i}^2$ parameters. After the
scale where  $m_{H_2}^2<0$, we define $m_{H_2}\ra -\sqrt{- m_{H_2}^2}$.
{}From these figures we conclude that the IGS symmetry can
break down radiatively naturally, provided that the two modular weights
are  different in  order to create a hierarchy for the two higgs mass parameters 
 at $M_{str}$, while the scale $m_{3/2}$ should not exceed $~(120-130)GeV$.

In this letter, we
 have analysed the modular invariance constraints on effective 
supergravity models with Intermediate Gauge Symmetry which 
usually arise in four dimensional string constructions. We find that
requirements for invariance of Yukawa terms in the superpotential
lead to specific relations for the modular weights of the 
massless spectrum of a particular model. 
Further constraints for the soft mass parameters are obtained,
in particular for the neutral higgs bosons associated with
the symmetry breaking. It is found that in particular cases
the Intermediate Gauge Symmetry breaks down to the standard
model radiatively.

\vspace*{1cm}

We would like to thank S. Dimopoulos and C. Kounnas for useful
suggestions and discussions
We also thank CERN - Theory division for kind hospitality
during the final stages of this work.

\newpage

{\bf Figure Captions}

\vspace*{1cm}

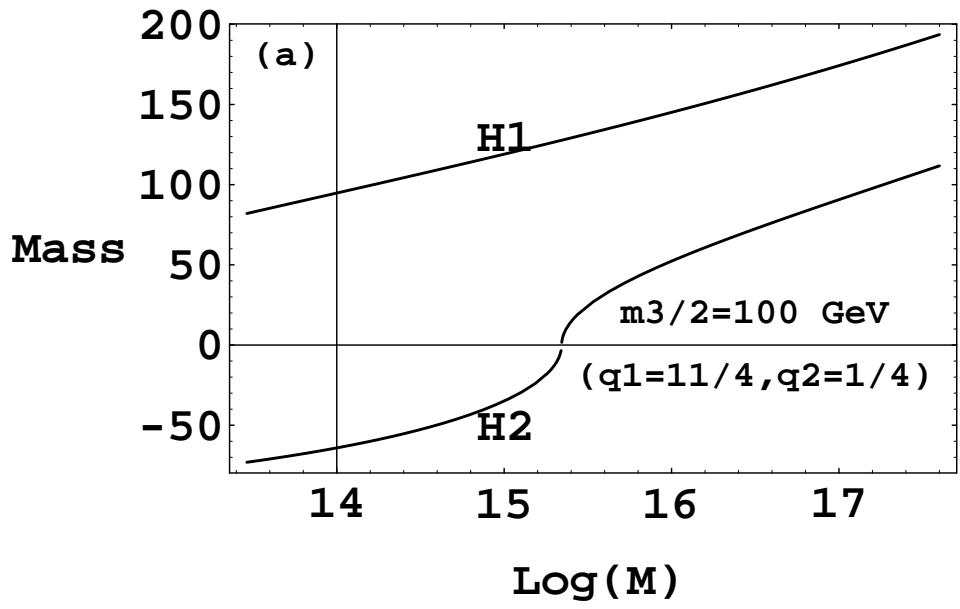
{\bf Figure 1:}$\;$ Plot of the higgs $mass$ - parameters  
for $m_{3/2}=100GeV$, {\it a) } $q_2=1/4 , q_1=11/4$  and 
{\it b) }$q_2=1, q_1=2$  as a function of the scale 
$\log_{10}M$.
For convenience, - treating properly the negative sign - we show 
here the $m_{H_1},m_{H_2}$, instead of the squared masses.

\vspace*{5mm}

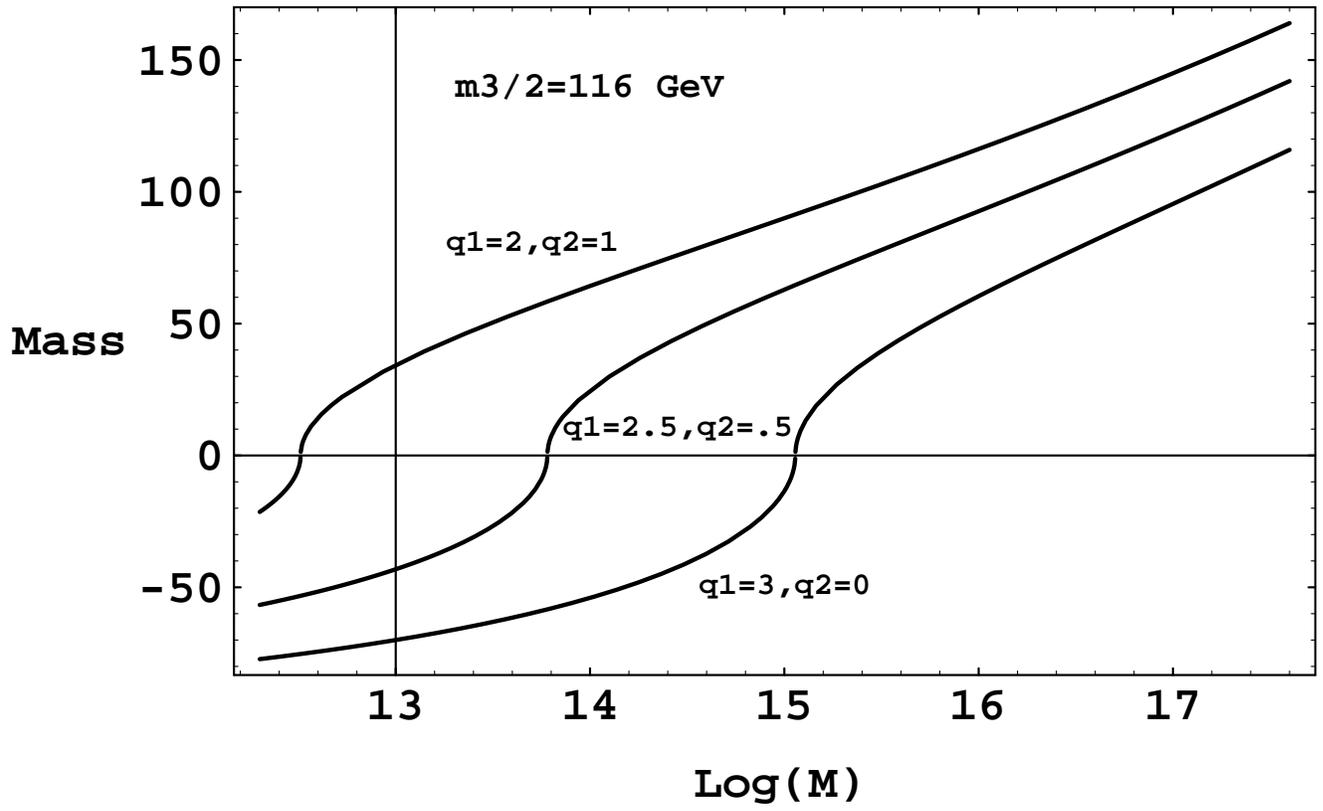
{\bf Figure 2:}$\;$ Plot of the $H_2$-higgs $mass$ - parameter  for
three $(q_1, q_2)$ - pairs and $m_{3/2}=116GeV$.
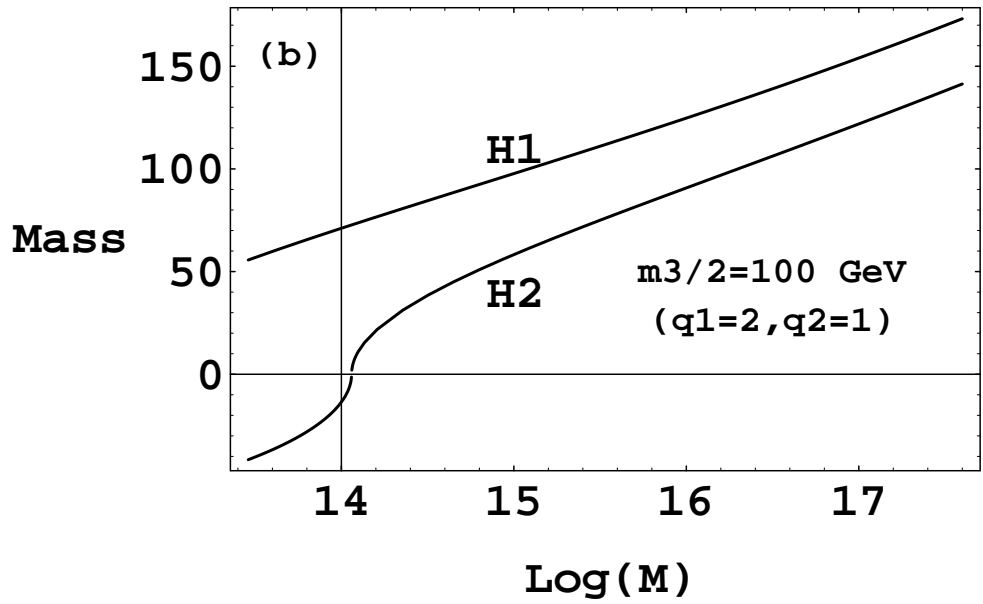
\end{document}